\documentclass[twoside]{article}
\usepackage{fleqn,espcrc2}

\usepackage{epsfig}

\newcommand{\AmS}{{\protect\the\textfont2
  A\kern-.1667em\lower.5ex\hbox{M}\kern-.125emS}}

\title{Baryon Masses in the $1/N_c$ Expansion}

\author{E. Jenkins\address[MCSD]{Department of Physics, 
        University of California, San Diego, \\ 
        9500 Gilman Drive, La Jolla, California 92093-0319 USA}
        \thanks{Supported by the Department of Energy under Contract No.
	DOE-FG03-97ER40546.}}
               
\begin{document}

\begin{abstract}
The masses of baryons and heavy quark baryons are studied {\it analytically}
in an expansion in $1/N_c$, $SU(3)$ flavor symmetry breaking, and heavy-quark
symmetry breaking.  The measured baryon masses are in striking agreement with
the $1/N_c$ hierarchy. 
\vspace{1pc}
\end{abstract}

\maketitle

\section{INTRODUCTION}

Large-$N_c$ spin-flavor symmetry has proven to be a useful symmetry for 
studying the properties of baryons in QCD~\cite{Jenk98}.  
Baryon observables can be analyzed
in an expansion involving a complete operator basis with each operator
transforming according to a definite representation under the unbroken 
spin $\otimes$ flavor subgroup and suppressed by a definite power of $1/N_c$.
From the $1/N_c$ expansion, it is possible to determine the symmetry structure
of $1/N_c$ corrections which account for deviations from the large-$N_c$
spin-flavor limit.  By including additional parameters to account for explicit
flavor symmetry breaking, one obtains a combined operator expansion in $1/N_c$ 
and flavor symmetry breaking.  The symmetry relations which occur when one
neglects individual operators in the combined $1/N_c$ and flavor-symmetry
breaking expansion are predicted to hold at a definite order in $1/N_c$ and
$SU(3)$ flavor symmetry breaking.  The hierarchy of symmetry relations predicted
by this procedure results in a nontrivial pattern for baryons in QCD since
$1/N_c = 1/3$ and the $SU(3)$ flavor symmetry breaking $\epsilon \equiv
m_s/\Lambda_\chi$ are comparable in magnitude, 
so the pattern of relations is not dominated
by either $1/N_c$ or $\epsilon$, and neither parameter 
can be neglected relative to the other.

The hierarchy of baryon mass relations predicted by the combined expansion in 
$1/N_c$ and flavor symmetry breaking has been obtained for the lowest-lying
baryon spin-flavor multiplet, which is the $\bf 56$ of $SU(6)$ for $N_c=3$
colors and $N_f =3$ light quark flavors.   This spin-flavor multiplet
contains the spin-$1/2$ octet and spin-$3/2$ decuplet baryon 
spin $\otimes$ flavor representations.  The mass hierarchy predicted for these
ground-state baryons is particularly predictive since the expansion extends to
relative order $1/N_c^3$ in the $1/N_c$ expansion and to third order in $SU(3)$
flavor symmetry breaking.  The mass spectrum of the spin-$1/2$ octet
and the spin-$3/2$ decuplet baryons is in striking accord with the $1/N_c$
hierarchy.  Suppression factors of $1/N_c =1/3$ are clearly evident in the data,
which provides the strongest single piece of evidence for the relevance of the
$1/N_c$ expansion for QCD baryons where $1/N_c =1/3$.  Even more importantly,
the $1/N_c$ expansion resolves the long-standing puzzle of why certain famous 
mass relations of the spin-$1/2$ octet and spin-$3/2$ decuplet baryons, such as
the Gell-Mann--Okubo formula, Gell-Mann's Equal Spacing Rule, the 
Okubo formula and the Coleman-Glashow relation, work so
extraordinarily well, which is to say, better than one would have expected 
based on flavor symmetry breaking suppression factors alone.  The point is that
the operators in the baryon mass expansion which are highly suppressed in
$SU(3)$ flavor symmetry breaking parameters also are suppressed by powers of
$1/N_c$, so that, generically, mass relations which are highly suppressed in
flavor are highly suppressed in $1/N_c$ as well.

This proceedings provides a brief description of progress that has
been made in understanding the mass spectrum of QCD baryons in an 
$\it analytic$ expansion in $1/N_c$ and flavor symmetry breaking.  A brief
description of the $1/N_c$ operator expansion for baryons is provided, with
the example of the operator expansion for the $I=0$ masses of the ground state 
baryons presented in detail.  Evidence of the $1/N_c$ mass hierarchy 
of the baryon octet and
decuplet is obvious for the $I=0$ baryon mass combinations, and 
recent experimental evidence supporting the predicted $1/N_c$ hierarchy in the
$I=1$ sector is noted for the isospin-breaking Coleman-Glashow mass splitting. 
Finally, the extension of the $1/N_c$ expansion to baryons containing a single
heavy quark in Heavy Quark Effective Theory (HQET) is summarized briefly, and
then applied to the charm and bottom baryon mass spectra.  Using the $1/N_c$
expansion, a prediction was made for the $\Xi_c^\prime$ mass valid to a
precision of a couple MeV, which has recently been confirmed experimentally.
           
\section{$1/N_c$ EXPANSION}

A contracted $SU(2 N_f)$ spin-flavor algebra for baryons arises in the 
large-$N_c$ limit.  For the simplest case of $N_f =2$ light quark flavors, 
the spin-flavor generators are given by baryon spin $J^i$, isospin $I^a$ 
and commuting generators $X^{ia}$.  
The contracted large-$N_c$ spin-flavor algebra is given by 
\begin{eqnarray}
&&\hspace{-20pt}\left[ X^{ia}, X^{jb} \right] =0, \nonumber\\
&&\hspace{-20pt}\left[ J^i, X^{ja} \right] = i \epsilon^{ijk} X^{ka},\ 
\left[ I^a, X^{ib} \right] = i \epsilon^{abc} X^{ic}, \qquad \\
&&\hspace{-20pt}\left[ J^i, J^j \right] = i \epsilon^{ijk} J^k,\ 
\left[ I^a, I^b \right] = i \epsilon^{abc} I^c, \nonumber
\end{eqnarray}
where the commuting generators $X^{ia}$ are obtained by rescaling the
axial vector baryon currents $A^{ia}$, which grow as ${\cal O}(N_c)$, by $1/N_c$
and taking the large-$N_c$ limit, which yields a finite well-defined result: 
\begin{equation}
X^{ia} \equiv \lim_{N_c \rightarrow \infty} {A^{ia} \over N_c}\ . 
\end{equation}
When working at finite $N_c$, it is useful to work with the operator
\begin{equation}
G^{ia} = \left( q^\dagger {{\sigma^i \tau^a} \over 4} q \right)
\end{equation}
of the large-$N_c$ quark model, and to use the uncontracted $SU(2N_f)$ algebra
of the quark model.  In this case, $G^{ia}/N_c$ has finite baryon matrix
elements equal to $X^{ia}$ in the large-$N_c$ limit.

The $1/N_c$ expansion of an arbitrary QCD $1$-body operator is given by
\begin{equation}
\bar q \, \Gamma q = \ N_c \ \sum_{n=0}^{N_c} \ {1 \over {N_c^n}}\ \ 
c_n\left({1 \over N_c} \right) \ {\cal O}_n \ ,
\end{equation}
where ${\cal O}_n$ is an $n$-body quark operator which equals an $n^{th}$ 
order polynomial
in the spin-flavor generators $J^i$, $T^a$, $G^{ia}$, and the operator 
coefficients $c_n(1/N_c) = {\cal O}\left(1 \right) + \cdots$ are unknown, 
but have a $1/N_c$ expansion beginning at order unity.  The baryon matrix
elements of the operators ${\cal O}_n$ are known, since the baryon matrix 
elements of the spin and flavor generators are known, as are the matrix elements
of $G^{ia}$.  (Only the leading ${\cal O}(1)$ matrix elements of the 
spin-flavor generators $X^{ia}$ are determined; the matrix elements of the
spin-flavor generators at
subleading orders are not, but any reasonable
choice will do.  The large-$N_c$ quark model generator $G^{ia}$ or the Skyrme
model collective coordinate $X_0^{ia}$ are two possible choices.)  The complete
set of independent
$n$-body operator products are classified.  There is a single $0$-body
operator $\bf 1$.  The independent $1$-body operators are the $SU(6)$ 
spin-flavor generators $J^i$, $T^a$ and $G^{ia}$, and $n$-body operators are
products of these $1$-body operators.  
Since not all $n$-body operator products are independent, operator identities 
are needed to reduce the operator products to an independent set.  The complete
set of operator identities is known.

\section{BARYON MASS HIERARCHY}

The baryon mass operator transforms as a spin singlet.  
$SU(3)$ flavor symmetry breaking proportional to the parameter $\epsilon$ 
transforms as the eighth component of an $SU(3)$ octet, so the baryon mass
operator expansion also contains $SU(3)$ flavor-symmetry breaking operators
which extend to third order in $SU(3)$ breaking and 
transform as the $\bf 8$, $\bf 27$ and $\bf 64$ representations of
$SU(3)$.  The baryon $I=0$ masses can be organized according to their $SU(3)$
representation,    
\begin{equation}
M = M^{\bf 1} + M^{\bf 8} + M^{\bf 27} + M^{\bf 64}\ ,
\end{equation}
with $1/N_c$ operator expansions
\begin{eqnarray}\label{m0}
&&\hspace{-22pt}M^{\bf 1} = N_c {\bf 1} + {1 \over N_c} J^2, \nonumber\\
&&\hspace{-22pt}M^{\bf 8} = T^8 + {1 \over N_c} \left\{ J^i, G^{i8} \right\} + {1 \over N_c^2}
\left\{ J^2, T^8 \right\}, \\
&&\hspace{-22pt}M^{\bf 27} = {1 \over N_c} \left\{ T^8, T^8 \right\} + {1 \over N_c^2} 
\left\{ T^8, \left\{ J^i, G^{i8} \right\} \right\}, \nonumber\\
&&\hspace{-22pt}M^{\bf 64} = 
{1 \over N_c^2} \left\{ T^8, \left\{ T^8, T^8 \right\} \right\}\ ,
\nonumber
\end{eqnarray}
where each operator in the $1/N_c$ expansion is understood to be accompanied 
by a coefficient, which has been suppressed in Eq.~(\ref{m0}) for simplicity.  
The above operator expansion consists of eight independent operators 
corresponding to the 
eight $I=0$ baryon masses $N$, $\Lambda$, $\Sigma$, $\Xi$, $\Delta$, 
$\Sigma^*$, $\Xi^*$ and $\Omega$.  Each mass operator occurs at a
definite order in $1/N_c$ and $\epsilon$ and corresponds to one specific
mass splitting, so the combined expansion in $1/N_c$ and $\epsilon$ predicts
a hierarchy of $I=0$ masses~\cite{Jenk95}.  A similar analysis can be performed for the $I=1$
and $I=2$ mass splittings of the octet and decuplet baryons.  The analysis
introduces two new parameters $\epsilon^\prime$ and $\epsilon^{\prime \prime}$
which are $I=1$ and $I=2$ flavor symmetry breaking parameters, respectively. 
The two parameters are expected to be similar in magnitude.

Figure~1 plots sixteen independent mass splittings of the spin-$1/2$ octet and
the spin-$3/2$ decuplet.
The plot shows, relative to the leading ${\cal O}(N_c)$ baryon mass,
$I=0$ mass splittings of 
order ${1 \over N_c^2}$, ${\epsilon \over N_c}$, 
${\epsilon \over N_c^2}$, ${\epsilon \over N_c^3}$, ${\epsilon^2 \over N_c^2}$, ${\epsilon^2 \over N_c^3}$
and ${\epsilon^3 \over N_c^3}$; $I=1$ mass splittings of order
${\epsilon^\prime \over N_c}$, ${\epsilon^\prime \over N_c}$,
${\epsilon^\prime \over N_c^3}$ or ${{\epsilon^\prime \epsilon} \over N_c^2}$, 
${{\epsilon^\prime \epsilon} \over N_c^2}$, 
${{\epsilon^\prime \epsilon} \over N_c^2}$, and
${{\epsilon^\prime \epsilon^2} \over N_c^3}$; and $I=2$ mass splittings of
order ${\epsilon^{\prime \prime} \over N_c^2}$, 
${\epsilon^{\prime \prime} \over N_c^2}$, and 
${{\epsilon^{\prime \prime} \epsilon} \over  N_c^3}$, respectively. 

The first seven points in Figure~1, the $I=0$ mass splittings, are in striking
agreement with the hierarchy predicted by the $1/N_c$ and $\epsilon$ expansions.
The singlet mass splitting of relative order $1/N_c^2$ is comparable in
magnitude to the octet mass splitting of relative order $\epsilon/N_c$.  The
three
flavor-octet mass splittings proportional to one power of $\epsilon$ are in
accord with the predicted
$1/N_c$ hierarchy of $1/N_c$, $1/N_c^2$, and $1/N_c^3$.
If one did not know
about the $1/N_c$ expansion and arbitrarily looked at three different
flavor-octet mass

\newpage

\begin{figure}
\epsfig{file=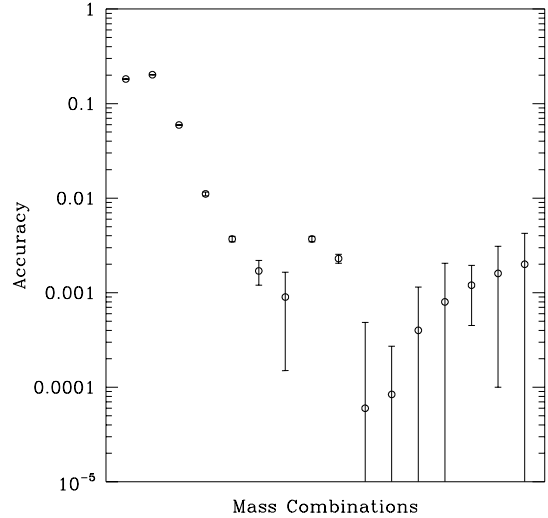,width=75mm}
\caption{Hierarchy of mass splittings for the lowest-lying baryon spin-flavor 
multiplet consisting of the spin-${1 \over 2}$ octet and
the spin-${3 \over 2}$ decuplet.  The experimental accuracy of the mass
combinations are plotted with $2.5\ \sigma$ error bars. 
}
\end{figure}

\noindent
splittings, they would generically all contain some portion of
the $1/N_c$ mass combination and therefore be of this relative magnitude. 
Only
when examining the linear combinations picked out by the $1/N_c$ expansion does
the $1/N_c$ hierarchy become apparent.   
The two 
flavor-${\bf 27}$ mass splittings are proportional to two powers of $\epsilon$, 
and occur
at relative orders $1/N_c^2$ and $1/N_c^3$ in the $1/N_c$ expansion.  These two
flavor-${\bf 27}$ mass splittings of the $1/N_c$ expansion are linear
combinations of the Gell-Mann--Okubo flavor-${\bf 27}$
mass splitting of the spin-$1/2$ baryon octet
\begin{equation}
{1 \over 4}\left( 2N -\Sigma - 3 \Lambda +2 \Xi \right) 
\end{equation}
and the flavor-${\bf 27}$ Equal Spacing Rule mass splitting of the spin-$3/2$
baryon decuplet  
\begin{equation}
{1 \over 7}\left(4 \Delta - 5 \Sigma^* -2\Xi^* + 3 \Omega \right) \ .
\end{equation}
Since these two famous mass splittings are linear combinations of the two 
$1/N_c$ mass
splittings, they are both suppressed by a factor of $1/N_c^2$ in addition 
to the
flavor suppression factor $\epsilon^2$.  Thus, the $1/N_c$ expansion predicts
that these flavor-${\bf 27}$ mass splittings are about an order of magnitude
smaller than expected from an analysis of flavor-symmetry breaking alone. 
The single flavor-${\bf 64}$ mass splitting
\begin{equation}
{1 \over 4} \left(\Delta - 3 \Sigma^* + 3 \Xi^* - \Omega \right)
\end{equation}
is third order in
$SU(3)$ flavor symmetry breaking $\epsilon^3$ and relative order $1/N_c^3$ 
in the $1/N_c$
expansion.  The experimental data clearly show that the mass splitting is
suppressed by a greater factor than expected from flavor-symmetry breaking
alone, and the observed suppression is consistent with the $1/N_c$ hierarchy.  A
better measurement of this splitting would reduce the experimental error bar and
test the $1/N_c^3$ prediction of the $1/N_c$ expansion.     

There also is clear evidence for the $1/N_c$ hierarchy in the $I=1$ mass
splittings~\cite{Jenk95,Jenk00}.  A new, very precise, measurement of
$\Xi^0 = 1314.82 \pm 0.06 \pm 0.2 \ {\rm MeV}$~\cite{Fant00} significantly 
reduces the errors on the $I=1$ mass combinations plotted in Fig.~1.  For the
first time, the Coleman-Glashow mass splitting
\begin{equation}
\left[ \left( p - n \right) - \left( \Sigma^+ - \Sigma^- \right) + 
\left( \Xi^0 - \Xi^- \right) \right] 
\end{equation}
is measured to be non-zero, though only at the $1\sigma$ level.
The Coleman-Glashow mass combination,
the eleventh point in Fig.~1, is predicted to be of relative order
${{\epsilon \epsilon^\prime} \over N_c^2}$, or suppressed by a factor 
of ${\epsilon \over N_c}$
relative to the leading $I=1$ mass splittings, the eighth and ninth points,
which are of relative order ${\epsilon^\prime \over N_c}$.  
A similar
suppression factor is predicted and observed for the $I=1$ mass splitting that
is plotted as the tenth point in Fig.~1.  The leading $I=2$ mass splitting,
point fourteen, is also in line with the prediction of the $1/N_c$ expansion.
Remaining points have large error bars.  Improved measurements of a number of
isospin mass splittings, particularly of the $\Delta$ baryons, would reduce
these errors and provide a further test of the $1/N_c$ hierarchy.   

\section{HEAVY QUARK BARYONS}

Large-$N_c$ baryons containing a single heavy quark respect a large-$N_c$
$SU(6)_\ell \times SU(4)_h$ spin-flavor symmetry~\cite{Jenk96} 
with light-quark generators $J_\ell^i$, $T_\ell^a$ and
$G_\ell^{ia}$ and heavy-quark generators 
\begin{eqnarray}
&&J_h^i = Q^\dagger {\sigma^i \over 2} Q = J_c^i + J_b^i, \nonumber\\
&&I_h^a = Q^\dagger {\tau^a \over 2} Q ,\\
&&G_h^{ia} = Q^\dagger {{\sigma^i \tau^a} \over 4} Q\ .\nonumber
\end{eqnarray}
Light-quark spin-flavor symmetry is valid in the $N_c \rightarrow \infty$
limit, whereas the heavy-quark spin-flavor symmetry is valid in the
large-$N_c$ limit and the heavy-quark limit of HQET.

The $1/N_c$ operator expansion for a baryon with $Qqq$ flavor quantum numbers
involves operator products which are at most $1$-body in the heavy-quark 
generators and at most $2$-body in the light-quark generators.
Light-quark generators are accompanied by a factor $1/N_c$, and heavy-quark 
generators are accompanied by a factor of $1/N_c$ and $\Lambda/m_Q$     
since the heavy-quark spin-flavor generators separately 
violate large-$N_c$ heavy-quark
spin-flavor symmetry and heavy-quark spin-flavor symmetry of HQET.
Heavy-quark flavor symmetry breaking for the $Q=c$ and $Q=b$ baryons enters
through the generators  
\begin{eqnarray}
&&I_h^3 = {1 \over 2} \left( N_{\rm charm} - N_{\rm bottom} \right), 
\nonumber\\
&&G_h^{i3} = {1 \over 2} \left( J_c^i - J_b^i \right), \nonumber
\end{eqnarray}
whereas light-quark $SU(3)$ flavor symmetry breaking enters through light-quark
generators which transform as the eighth or third component of an $SU(3)$ octet.
The $1/N_c$ mass hierarchy can be analyzed separately for charm- and
bottom-quark baryons using only heavy-quark spin symmetry, or together using
heavy-quark spin-flavor symmetry.  The joint expansion leads to  
relations between $Q=b$ and $Q=c$ mass splittings such as
\begin{equation}
{1 \over 3} \left( \Sigma_b + 2 \Sigma_b^* \right) - \Lambda_b
= {1 \over 3}\left(\Sigma_c + 2 \Sigma_c^* \right) - \Lambda_c\ .
\end{equation}
Heavy quark baryons and baryons containing no heavy quarks also can be analyzed
in a joint expansion which involves the heavy-quark number operator
\begin{equation}
N_h = Q^\dagger Q = N_{\rm charm} + N_{\rm bottom} \ .
\end{equation}
This analysis leads to relations between heavy-quark baryon mass splittings and
mass splittings of the spin-$1/2$ octet and the spin-$3/2$ decuplet, such as
\begin{equation}
{ 1 \over 3} \left( \Sigma_Q + 2 \Sigma_Q^* \right) - \Lambda_Q = {2 \over 3}
\left( \Delta - N \right)\ .
\end{equation}

The $1/N_c$ expansion for heavy-quark baryon masses successfully predicted the
$\Xi_c^\prime$ mass: the predicted value of
$\Xi_c^\prime = 2580.8 \pm 2.1 \ {\rm MeV}$~\cite{Jenk97} is in 
excellent agreement with the
subsequent measured value of $\Xi_c^\prime = 2576.5 \pm 2.3 \ {\rm MeV}$.
With this measurement, all of the lowest-lying charm baryon masses are measured
except for $\Omega_c^*$.  The $\Omega_c^*$ mass can be determined in terms of 
the other measured charm baryon masses using the mass relation
\begin{equation}
{1 \over 4} \left[ \left(\Sigma_c^* - \Sigma_c \right) - 2 \left(\Xi_c^* -
\Xi_c^\prime \right) + \left( \Omega_c^* - \Omega_c \right) \right] \ ,
\end{equation}
which is predicted to hold at the sub-MeV level using the $1/N_c$ expansion, and
therefore is extremely accurate and
can be taken to be exact for most purposes.  Presently, the extraction of the
$\Omega_c^*$ mass using this mass relation is dominated by experimental
uncertainties in the $\Omega_c$ mass, whose PDG value of 
$\Omega_c = 2704 \pm 4 \ {\rm MeV}$ is $2.5 \sigma$ away from the new CLEO
measurement of $\Omega_c= 2694.6 \pm 3.5 \ {\rm MeV}$.  These two experimental
values yield central values for the 
$\Omega_c^*$ of 2768 and 2777 MeV, respectively.  Finally, all of the $Q=b$ baryons
(given the measured $\Lambda_b$ mass) can be predicted from the charm baryon
masses to a precision of about $10$ MeV, which is set by the accuracy of the
worst $Q=c$ and $Q=b$ mass relation.  Observation of some of the $Q=b$ baryons
will allow a determination of the remaining unobserved $Q=b$ baryon masses to
greater precision.

\section{CONCLUSIONS}

The $1/N_c$ expansion of QCD has provided an understanding of the mass
spectrum of the spin-$1/2$ octet and spin-$3/2$ decuplet baryons.
The mass hierarchy of
baryons is well-described by an expansion
in $1/N_c$ about the large-$N_c$ spin-flavor limit and an expansion in $SU(3)$
flavor symmetry breaking about the $SU(3)$ flavor symmetry limit.  
$1/N_c$ suppression factors are necessary to understand the relative magnitudes
of the various baryon mass splittings of the baryon octet and decuplet.
In the case of baryons containing a single heavy quark in HQET, there is an
additional expansion parameter $\Lambda/m_Q$ which describes corrections  
to the heavy-quark spin-flavor symmetry limit.  The hierarchy of heavy-quark
baryon masses predicted from the violation of large-$N_c$ light- and 
heavy-quark spin-flavor symmetries is sufficiently restrictive that it can be
used to predict heavy-quark baryon masses at a level which is interesting
experimentally.  The combined $1/N_c$ and flavor-symmetry breaking expansions
successfully predicted the $\Xi_c^\prime$ mass to an accuracy of
several MeV.  Additional predictions for charm and bottom baryon 
masses can and will be tested in the future.  

\

\end{document}